\documentclass[reprint,nopreprintnumbers,amsmath,amssymb,superscriptaddress,aps,prc,lengthcheck]{revtex4-1}
\usepackage{natbib}
\usepackage{graphicx}
\usepackage{subfigure}
\usepackage{bm}

\makeatletter
\newcommand\erfc{\mathop{\operator@font erfc}\nolimits}
\def\slashchar#1{\setbox0=\hbox{$#1$}
   \dimen0=\wd0 \setbox1=\hbox{/} \dimen1=\wd1
   \ifdim\dimen0>\dimen1 \rlap{\hbox to \dimen0{\hfil/\hfil}} #1
   \else  \rlap{\hbox to \dimen1{\hfil$#1$\hfil}} / \fi}

\makeatother

\begin{document}
\title{Directed  flow  in ultrarelativistic heavy-ion collisions}
\author{Piotr Bo\.zek}
\email{Piotr.Bozek@ifj.edu.pl}
\affiliation{The H. Niewodnicza\'nski Institute of Nuclear Physics,
PL-31342 Krak\'ow, Poland} \affiliation{
Institute of Physics, Rzesz\'ow University, 
PL-35959 Rzesz\'ow, Poland}
\author{Iwona Wyskiel}
\affiliation{The H. Niewodnicza\'nski Institute of Nuclear Physics,
PL-31342 Krak\'ow, Poland}
\date{\today}

\begin{abstract}
We study the generation of the directed 
flow in the hydrodynamic expansion of  hot matter formed
 in ultrarelativistic heavy-ion collisions 
at $\sqrt{s}=200$GeV. The experimentally observed 
negative directed flow in a wide range of central pseudorapidities is
reproduced, assuming that the  fireball is tilted away from the collision 
axis. The tilt of the source is consistent with a preferential emission in
the forward/backward hemisphere from forward/backward participating nucleons.
The model reproduces the experimentally observed scaling of the  directed flow
when going from 
Au-Au to Cu-Cu systems. 
\end{abstract}



\maketitle

\section{Introduction}

The appearance of the transverse, azimuthally asymmetric flow is 
one of the key observations in the physics of relativistic 
heavy-ions \cite{Arsene:2004fa,*Back:2004je,*Adams:2005dq,*Adcox:2004mh}. 
It proves that a collectively expanding, strongly interacting medium is formed
in the course of the reaction. A number of observables have been studied, 
both in experiments, and in model calculations, in order to unravel the 
 properties of
the dense hot matter created in the collisions. The production of particles 
with soft momenta can be interpreted as a thermal emission of
 particles from fluid elements moving with some collective velocity field
 \cite{Schnedermann:1993ws,*Kolb:2000sd}. Relativistic hydrodynamics
describes quantitatively the development of the 
collective velocity from pressure gradients in the fireball
 \cite{Teaney:2000cw,*Kolb:2003dz,*Hama:2005dz,*Huovinen:2006jp,*Hirano:2005xf,*Broniowski:2008vp,Hirano:2002ds,Bozek:2009ty}.

For non-central collisions the interaction region is azimuthally asymmetric 
and, as a result of the collective
 expansion of matter,
 azimuthally asymmetric 
emission of particles takes place. The effect can be 
quantified in terms of Fourier coefficients in the expansion of the
 measured particle spectra 
\begin{eqnarray}
\frac{dN}{d^2p_\perp d\eta}&=&\frac{dN}{2\pi p_\perp dp_\perp d\eta}\left(1 
+2 v_1 \cos( \phi) \right. \nonumber \\ 
&+& \left. 2 v_2 \cos(2\phi)+\dots  \right) \ .
\end{eqnarray}
The elliptic flow coefficient $v_2$   is known
 to be a very sensitive probe of the pressure in the system at  early 
stages \cite{Ollitrault:1992bk}.

Directed flow, quantified by the coefficient $v_1$ is also measured at
 energies available at the 
BNL Relativistic Heavy Ion Collider 
 (RHIC).  The coefficient $v_1$ of the directed flow is zero at 
zero rapidity
for collisions of symmetric nuclei, but it increases when moving to 
forward or backward 
pseudorapidities $\eta$. Its size and sign has been the subject of many studies 
at lower energies, where it is dominated by nucleon flow 
\cite{Herrmann:1999wu,*Wetzler:2002fi,*Voloshin:2002wa,*Lisa:2000ip}. At RHIC energies the
spectator nucleons have positive directed flow ($v_1>0$ for $\eta>0$),
resulting from 
the deflection of the spectators during the collision. On the other hand, 
 matter in the fireball shows a significant negative (anti-flow) component 
for pseudorapidities 
$-4<\eta<4$ and for centralities $c=0-80\%$, both for Cu-Cu 
and Au-Au collisions \cite{Back:2005pc,Adams:2005ca,Abelev:2008jga}. 
 A striking characteristic of the measured 
directed flow is the large negative value of $v_1$ even at the highest
 energy $\sqrt{s}=200$GeV. Another important observation is the scaling
 of the measured directed flow with the size of the system. The coefficient 
$v_1$ is the same for both systems (Au-Au or Cu-Cu),
for the same centrality. Whereas,
 a scaling of $v_2$ with the density 
of the fireball has been observed \cite{Voloshin:2007af}. 

Transport models of nuclear reactions,
 describe the directed flow at lower energies, 
but generally underpredict the amount of anti-flow at RHIC energies 
at central rapidities
 \cite{Bleicher:2000sx,*Chen:2009xc}. On the other hand, 
these calculations predict a large negative flow at large rapidities.
 Some calculations \cite{Burau:2004ev}
 predict significant anti-flow 
around central rapidities, larger than observed,  but yield a 
positive flow for 
$|\eta|>3$, unlike observed in experiments. It has been noticed that the
 appearance of negative directed flow 
around  central rapidities could be an effect of the softening of
 the equation of state  \cite{Csernai:1999nf}. This effect 
is called the third flow component. Hydrodynamic calculations incorporating
 such effects yield a negative elliptic flow at central rapidities and 
  a positive directed flow at larger pseudorapidities, unlike  the 
experimental data. A hydrodynamic calculation with initial conditions from 
a microscopic model gives the correct sign, but a too small magnitude 
of the directed flow for central rapidities  \cite{Andrade:2008fa}.

There are two effects leading to a negative directed flow in the models. 
The first one is  the shadowing of the fireball matter 
by the spectators, which  can
 give a substantial negative $v_1$ for $|\eta|>4$.
The second one is the build up of the flow away from the 
collision axis due to a tilt of the source. 
The description of the  magnitude of the directed flow at different 
centralities for central rapidities requires, a  tilt of the source of 
the right magnitude as 
function 
of the impact parameter, and a sufficient  amount of collectivity to 
generate the flow. 
We consider two initial conditions for the hot source in the hydrodynamic 
evolution. The first one is the most commonly used initial conditions 
in $3+1$-dimensional ($3+1$D) calculations
incorporating a Bjorken flow in the longitudinal direction and a 
shift in space-time 
rapidity due to the local imbalance of the momentum \cite{Hirano:2002ds}.
The second one assumes that the initial density results from a superposition 
of the energy density radiated by the color sources in the target 
and the projectile.  
The preferred emission in the pseudorapidity hemisphere of 
the emitting charge results in a tilt of the source for non-central collisions.
We show that the second choice leads to a satisfactory
 description of the directed 
flow generated in heavy ion collisions at $\sqrt{s}=200$GeV
 for central rapidities.

\section{Initial conditions and early flow}

Hydrodynamic evolution in $3+1$D at RHIC energies is performed in the
 proper time $\tau=\sqrt{t^2-z^2}$. The densities are defined as 
functions of the 
transverse plane ($x$-$y$) coordinates and the space-time rapidity
 $\eta_\parallel=\frac{1}{2} \log\left((t+z)/(t-z)\right)$.
The hydrodynamic model requires some initial 
 density and initial flow profile to be chosen at the initial
  time $\tau_0$.
Although some guidance from microscopic models of elementary
 collisions is possible in the choice of the initial conditions, there
 is still a vast choice of initial energy density profiles  that are used 
in the simulations. Also, most of the calculations assume 
a Bjorken initial flow in the longitudinal direction
and no initial transverse flow 
\begin{equation}
u^\mu(\tau_0,x, y,\eta_\parallel)=(\cosh \eta_\parallel
 , 0 , 0, \sinh \eta_\parallel) \ .
\label{eq:bj}
\end{equation}

In the hydrodynamic evolution the generation of the left-right 
asymmetry of the flow in  the two (forward/backward) 
halves  of the reaction plane 
 requires the presence of an asymmetry in the initial distributions.
For non-central collisions the azimuthal asymmetry of the interaction region
 results in a non-zero initial eccentricity in the transverse plane 
that gives rise to   the collective elliptic flow.  For collisions of symmetric 
nuclei and neglecting the fluctuations, the odd components of the decomposition 
in Fourier coefficients vanish at space-time rapidity zero. 
At forward and
 backward rapidities an imbalance between the contributions from  the target 
and the projectile to the initial source can result in a left-right deformation 
of the source in the transverse plane and/or in an asymmetric initial flow, 
that could generate collective 
directed flow.

The observation of non-zero directed flow implies, that the symmetry in the 
reaction 
plane is indeed  broken, either in the initial flow or in the 
initial density, or both.
We tried to reproduce the observed directed flow assuming an 
asymmetric initial flow,
different from the Bjorken one (\ref{eq:bj}), but without success.
 Therefore
 in the following we assume a Bjorken initial flow and study the 
effect of asymmetric
 initial densities on the directed flow. 

  Starting with a factorized initial energy
 density 
in the transverse plane and in the space-time rapidity 
\begin{equation}
\epsilon(\eta_\parallel,x,y)= \rho(x,y) f(\eta_\parallel)
\label{eq:sym}
\end{equation}
 the subsequent evolution remains symmetric 
with respect to
 the $\eta_\parallel$ axis in the reaction plane ($\eta_\parallel$-$x$),
 with the 
consequence that the directed flow is exactly zero. Factorized 
initial conditions 
in the Glauber Model imply that all the participant nucleons or binary
 collisions
contribute in a similar way to the total density, with a longitudinal profile 
proportional to $f(\eta_\parallel)$.

\begin{figure}
\includegraphics[width=.4\textwidth]{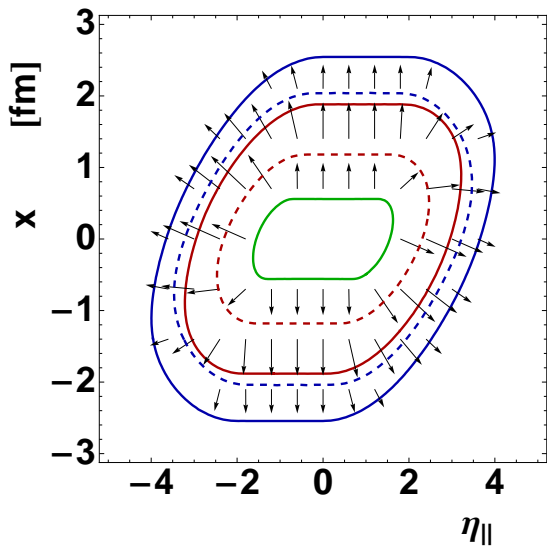}
\caption{(Color online) 
Contour  plot of the initial pressure $p(\eta,x,y=0)$ in the fireball 
for the  shifted densities (Eq. \ref{eq:nwdens}).
 Solid lines correspond to the 
 pressure of $9$, $3$ and $1$GeV/fm$^3$ for Au-Au collisions (impact parameter 
$b=11$fm) and dashed lines 
to the pressure of $3$ and $1$GeV/fm$^3$ for Cu-Cu collisions ($b=7.6$fm). 
The 
arrows represent the gradient $(-\partial_\eta p/\tau_0,-\partial_x p)$ 
for Au-Au collisions
in arbitrary units. }
\label{fig:indens1}
\end{figure}

Modifications of the symmetric distribution (\ref{eq:sym}) could happen during 
 the formation of the initial thermalized state.
Momentum imbalance between left and right going participants at
 a given point in the transverse plane
results in a nonzero total momentum of the matter.
The longitudinal distribution is shifted in space-time rapidity
 \cite{Hirano:2001yi,Hirano:2002ds} by the value of the center of mass 
rapidity of the fluid
\begin{equation}
\eta_{sh}=\frac{1}{2}\log\left(\frac{N_++N_-+v_N(N_+-N_-)}
{N_++N_--v_N(N_+-N_-)}\right) \ ,
\label{eq:etash}
\end{equation}
where $N_+$ and $N_-$ are the densities of participants from the two nuclei and
 $v_N$ is the velocity of the incident nuclei.
\begin{eqnarray}
N_+(x,y)& =& T(x-b/2,y)\left(1-\exp(- \frac{\sigma T(x+b/2,y)}{A})\right) \nonumber \\
N_-(x,y)& =& T(x+b/2,y)\left(1-\exp(- \frac{\sigma T(x-b/2,y)}{A})\right)  \ ,
\end{eqnarray}
where $\sigma$ is the cross section,
\begin{equation}
T(x,y) =\int dz \rho(x,y,z) 
\end{equation}
is the thickness function calculated from the 
Woods-Saxon density of colliding nuclei
\begin{equation}
\rho(x,y,z)=\frac{\rho_0}{1+\exp\left((\sqrt{x^2+y^2+z^2}-R_A)/a\right)} \ .
\end{equation}
We use the same 
parameters as in Ref. \cite{Bozek:2009ty}, where a satisfactory 
description of spectra and femtoscopy data for Au-Au collisions
 at $\sqrt{s}=200$GeV has been obtained.
The initial energy density takes the form \cite{Hirano:2001yi,Hirano:2002ds}
\begin{eqnarray}
\epsilon(\tau_0)&=& \epsilon_0  f(\eta_\parallel-\eta_{sh})\left[
 \left( N_+(x,y)+ N_-(x,y)\right)
(1-\alpha) \right.\nonumber \\
&+& \left. 2 \alpha N_{bin}(x,y)\right] /N_0 \ .
\label{eq:nwdens}
\end{eqnarray}
The relative contribution from binary collisions is $\alpha=0.145$, with
$N_{bin}(x,y)=\sigma T(x-b/2,y)T(x+b/2,y)$. 
In the following we call these initial conditions shifted initial conditions
(Fig. \ref{fig:indens1}). 

The form of the initial energy density profile $f(\eta_\parallel)$ 
is adjusted to reproduce the measured
 charged particle distribution in pseudorapidity. The resulting
width of the initial 
distribution depends 
on the chosen equation of state, initial time $\tau_0$ and shear viscosity 
\cite{Satarov:2006iw,*Bozek:2007qt,Bozek:2009ty}. In the following
 we use ideal fluid hydrodynamics with $\tau_0=0.25$fm/c and a realistic, 
hard equation of state \cite{Chojnacki:2007jc}, which requires
\begin{equation}
f(\eta_\parallel)=
\exp\left(-\frac{(\eta_\parallel-\eta_0)^2}{2\sigma_\eta^2}\theta(|\eta_\parallel|-\eta_0)
\right)
\label{eq:etaprofile}
\end{equation}
with a plateau of width $2\eta_0=2.0$ and $\sigma_\eta=1.3$.

\begin{figure}
\includegraphics[width=.48\textwidth]{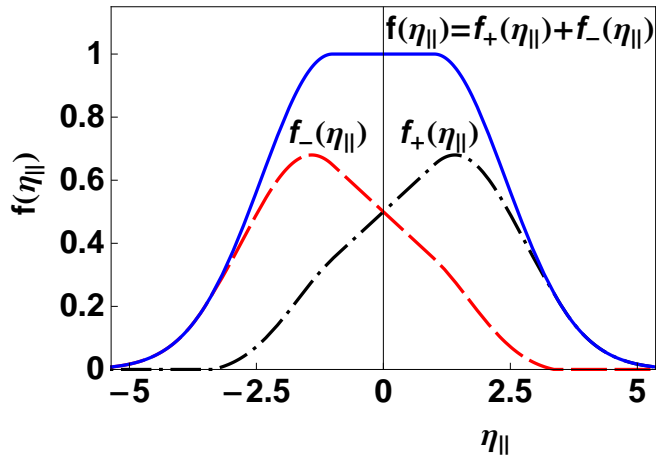}
\caption{(Color online) Initial profile in the longitudinal
 (space-time rapidity) direction. The symmetric function $f(\eta_\parallel)$ 
is composed  from two contributions $f_{+}$ and $f_{-}$ 
representing the emission from forward and backward going 
participant nucleons.}
\label{fig:ff}
\end{figure}

\begin{figure}
\includegraphics[width=.4\textwidth]{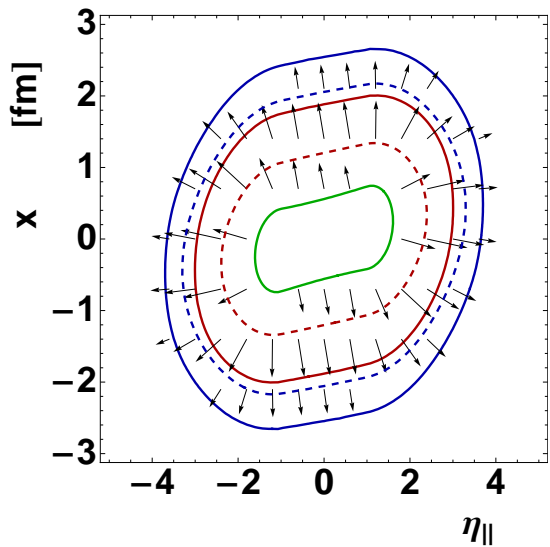}
\caption{(Color online) Same as in Fig. \ref{fig:indens1} but for tilted 
initial conditions (Eq. \ref{eq:tilt}).}
\label{fig:indens2}
\end{figure}

A different type of initial conditions studied in this work assumes a preferred
 emission from participating nucleons in the same hemisphere. Instead 
of a symmetric distribution of matter in space-time rapidity given by the 
function $f(\eta_\parallel)$ in Eq. (\ref{eq:sym}), we assume that the 
deposited energy depends on the rapidity of the emitting participating nucleon.
Such a distribution 
 depending on the rapidity difference
 between the  emitting charge and the emitted gluon is assumed in 
some phenomenological models \cite{Bialas:2004kt,*Fialkowski:2004wh,*Brodsky:1977de,*Adil:2005qn,*Adil:2005bb}. However,
there is no 
direct measurement of the contribution to soft particle production
 from a single forward or backward moving charge. A phenomenological 
analysis is possible, by comparing multiplicity distributions in pseudorapidity 
 for different  asymmetric systems   or by studying multiplicity 
correlations in different pseudorapidity intervals.
These studies  indicate that a preferred emission for 
rapidities close to the rapidity of the participating charge 
occurs \cite{Bialas:2004su,Bzdak:2009xq,Gazdzicki:2005rr,Bzdak:2009dr}.
In the wounded nucleon model of nuclear collisions,  such
 correlations can be understood as due to a specific distribution of soft 
particles produced by each participant nucleon. Nucleons from the projectile
 (with positive rapidity $y_B=\ln(\sqrt{s}/m_N)>0$) emit
 more particles in the forward ($\eta>0$) than in the backward hemisphere.
The form of the 
extracted charged particle distribution can be approximated by the function
\begin{equation}
f_F(\eta)=\frac{\eta+\eta_m}{2\eta_m} \ ,
\end{equation}
in the interval $[-\eta_m,\eta_m]$, where $\eta_m=y_b-\eta_s$ 
defines 
the range of rapidity correlations, at $\sqrt{s}=200$GeV 
it is $\eta_m \simeq y_b-2\simeq 3.36$. The origin of the shift in rapidity 
$\eta_s\simeq 2$ is not understood \cite{Bialas:2004su,Bzdak:2009xq}.
 For practical
 purposes we can treat it as a phenomenological parameter. Particle 
production in the remaining pseudorapidity  intervals
 close to the fragmentation regions $[\eta_m,y_b]$ and $[-y_b,-\eta_m]$ cannot
 be reliably described in a hydrodynamic model anyway. Within the 
framework of relativistic hydrodynamics,  we are interested in describing the 
 main characteristics of the soft part of particle spectra  in the 
central region $-3.5<\eta<3.5$ and, in
 particular,   the 
directed flow.
There is another reason why  the phenomenological estimates of 
the emission of particles from participant nucleons  
\cite{Bialas:2004su,Bzdak:2009xq} cannot be directly translated into 
the  initial 
conditions for hydrodynamics, that we are interested in. Refs. 
\cite{Bialas:2004su,Bzdak:2009xq} study particle distributions 
and correlations in the final state, whereas, we know that in realistic
 hydrodynamic simulations the matter distribution in space-time rapidity 
evolves during the expansion of the fireball 
\cite{Satarov:2006iw,Bozek:2009ty}, also statistical 
emission broadens the distribution in pseudorapidity.
It means that the initial profile $f(\eta_\parallel)$ is significantly 
narrower than the final charged particle distribution $\frac{dN}{d\eta}$.  
The correlation functions in pseudorapidity \cite{Bzdak:2009xq} 
can be modified due to the 
longitudinal transport and the generation of  transverse flow as well.
We propose as a phenomenological  ansatz,  inspired by the observations  in
 Refs. \cite{Bialas:2004su,Bzdak:2009xq,Gazdzicki:2005rr,Bzdak:2009dr},
that the initial energy density of matter produced by a single 
participant nucleon of rapidity $y_b$ is proportional to 
\begin{equation}
f_{+}(\eta_\parallel)= f(\eta_\parallel) f_F(\eta_\parallel)
\end{equation}
where $f(\eta_\parallel)$ is the initial longitudinal profile 
(\ref{eq:etaprofile}) fitted to
 reproduce $\frac{dN}{d\eta}$ and
\begin{equation}
f_F(\eta_\parallel)=\begin{cases} 0 & \eta_\parallel< -\eta_m\\
\frac{\eta_\parallel+\eta_m}{2\eta_m}   & -\eta_m \le \eta_\parallel \le \eta_m \\
1 & \eta_m<\eta_\parallel
\end{cases}
\end{equation}
The initial energy density of the fireball is constructed as a sum of three 
terms 
originating from the forward or backward moving participant nucleons and from 
the binary collisions that are assumed to contribute in a symmetric way
\begin{eqnarray}
\epsilon(\tau_0)&=& \epsilon_0 \left[2
 \left( N_+(x,y)f_+(\eta_\parallel)+ N_-(x,y)f_-(\eta_\parallel)\right)
(1-\alpha) \right.\nonumber \\
&+& \left. 2 \alpha N_{bin}(x,y) f(\eta_\parallel)\right] /N_0 \ .
\label{eq:tilt}
\end{eqnarray}
The net result of the difference between forward and backward emission is a 
tilt of the 
source in the $x$-$\eta_\parallel$ plane (Fig. \ref{fig:indens2}). 
This breaks the symmetry in the longitudinal direction and generates nonzero
 directed flow in the expansion.

Hydrodynamic equations in $3+1$D 
\begin{equation}
\partial_\mu T^{\mu\nu}=0
\end{equation}
constitute four independent equations, 
which, together with the equation of state, determine the evolution of the 
energy density $\epsilon$, of the pressure $p$, and of  three independent
 components on the fluid velocity.
The fluid four velocity can be written in the form
\begin{equation}
u^\mu=(\gamma \cosh Y, u_x, u_y, \gamma \sinh Y) \ ,
\end{equation}
$u_x$ and $u_y$ are the components of the transverse velocity, 
$\gamma=\sqrt{1+u_x^2+u_y^2}$ and $Y=\frac{1}{2} \ln\left(\frac{1+v_z}{1-v_z}
\right)$ is the fluid rapidity. The densities are functions of the proper 
 time $\tau$, the space-time rapidity $\eta_\parallel$ and  the transverse 
coordinates $x$, $y$. The equations in the expanded form can be found in 
\cite{Bozek:2009ty}. At early times, the velocities on
 the right hand side of the equations 
can be approximated by the initial velocities
$u_x=0$, $u_y$=0, $Y=\eta_\parallel$. The two acceleration equations
 for the velocity components in the
 reaction plane take the form
\begin{eqnarray}
\partial_\tau u_x = - \frac{1}{\epsilon+p} \partial_x p \nonumber \\
\partial_{\tau} Y=  - \frac{1}{\tau(\epsilon+p)} \partial_{\eta_\parallel}p \ .
\label{eq:acce}
\end{eqnarray}
In $3+1$D hydrodynamic evolution, the lack of Bjorken invariance results in
 a nonzero longitudinal acceleration. The fluid rapidity
 $Y$ becomes larger than the space-time rapidity $\eta_\parallel$. 
In Figs. \ref{fig:indens1} and \ref{fig:indens2} are shown 
the vector fields of the initial pressure 
gradients ($\frac{1}{\tau_0}\partial_{\eta_\parallel} p,\  \partial_x p$). For 
the shifted initial conditions the gradient in the central region of the 
fireball
is  in the transverse direction, and
 mainly transverse flow is generated in the early stage.
This is due to the existence of an approximate Bjorken plateau 
for central rapidities in
the initial stage. 
The situation is different for tilted initial conditions 
(Fig. \ref{fig:indens2}); the acceleration in the tilted source is 
anti-correlated in the transverse $x$ and longitudinal $\eta_\parallel$
 directions. The matter that is accelerated to positive 
rapidities is preferably 
accelerated in the negative $x$ direction. 
In the same Figures are shown the lines of constant pressure  for the 
initial fireball created in Cu-Cu collisions for the same centrality. 
The deformation for the shifted fireball (Fig. \ref{fig:indens1}), or the tilt 
for the tilted fireball (Fig. \ref{fig:indens2}), in the smaller system 
is very similar as in the larger system. This generates a similar
 directed flow  in the two systems irrespective of their sizes.

\section{Results}

\begin{figure}
\includegraphics[width=.48\textwidth]{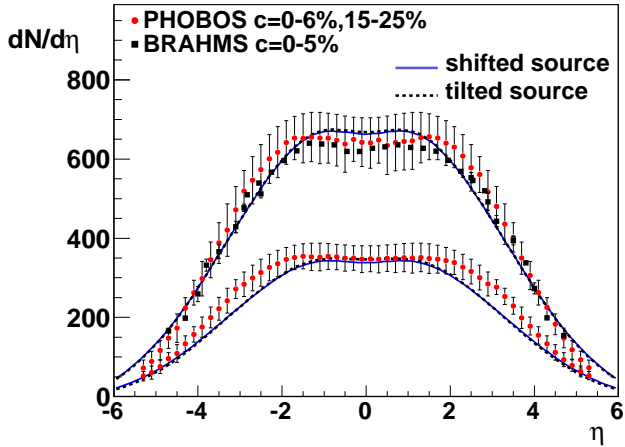}
\caption{(Color online) 
 Pseudorapidity distribution 
of charged particles for centrality classes $0-6\%$,  $15-25\%$, 
calculated for the shifted and tilted initial conditions
(solid and dashed 
lines respectively) compared to PHOBOS Collab. data (dots) 
\cite{Back:2002wb}. The squares represent the BRAHMS Collab. 
data for centrality $0-5\%$ 
  \cite{Bearden:2001qq}. }
\label{fig:dndeta}
\end{figure}

\begin{figure}
\includegraphics[width=.48\textwidth]{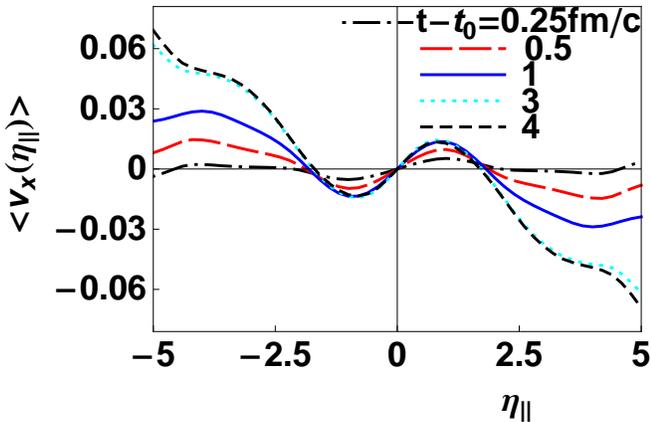}
\caption{(Color online) 
Average flow in the $x$ direction as function of space-time rapidity 
for different evolution times for the Hirano-Tsuda shifted initial densities 
(Eq. \ref{eq:nwdens}). }
\label{fig:v1timeht}
\end{figure}

\begin{figure}
\includegraphics[width=.48\textwidth]{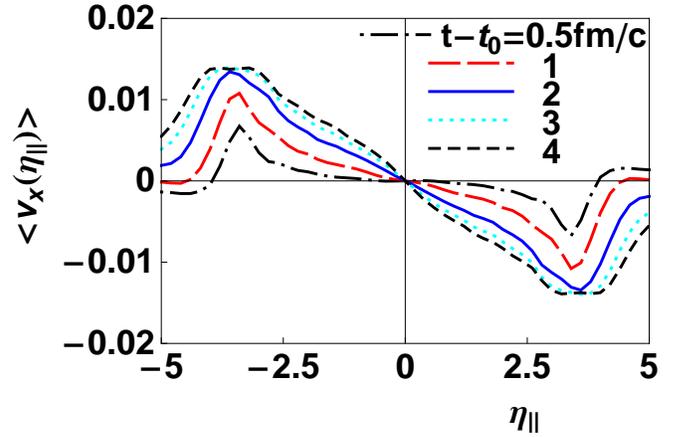}
\caption{(Color online) Same as in Fig. \ref{fig:v1timeht} but for tilted 
initial conditions (Eq. \ref{eq:tilt}).}
\label{fig:v1time}
\end{figure}

The hydrodynamic equations are solved numerically for the two sets
 of initial conditions, the shifted initial distributions (\ref{eq:nwdens})
and the tilted initial conditions (\ref{eq:tilt}). The 
parameter $\epsilon_0$ is chosen to reproduce  
particle spectra and multiplicities in 
central collisions. 
We use $\epsilon_0=107$GeV/fm$^{-3}$ and $\epsilon_0=65$GeV/fm$^{-3}$ 
for Au-Au and Cu-Cu collision respectively, and a freeze-out temperature 
of $T_F=150$MeV. This  gives a satisfactory description of the spectra 
in  collisions up to centralities of $50\%$ \cite{Bozek:2009ty}.

 The distribution of charged particles
 in pseudorapidity is shown in Fig. \ref{fig:dndeta}.  The results 
 obtained from 
the two initial conditions are almost indistinguishable on the plot.
 Both initial conditions lead 
to similar results for  transverse momentum spectra of particles, 
interferometry radii, and elliptic flow, as well. The comparison with 
experimental data can be found in Ref. \cite{Bozek:2009ty},
 giving satisfactory results.
The calculated elliptic flow  overshoots 
the experimental data since we do not take  viscosity effects into account
 \cite{Bozek:2009mz}.
The parameters of the 
 initial profile in space-time rapidity $f(\eta_\parallel)$ are adjusted
 to reproduce, as closely as possible, the experimental results on
 pseudorapidity distributions, and transverse momentum spectra at non-zero
rapidities \cite{Bozek:2009ty}.   Changing the width $\eta_0$
of the plateau in the initial profile $f(\eta_\parallel)$, within a 
range compatible with the observed pseudorapidity distributions, does 
not change
the results for the directed flow. Only  taking the unrealistic
value $\eta_0=0$ for shifted 
initial conditions,  causes the wiggle of positive $v_1$ 
disappear at central pseudorapidities.

By breaking the symmetry 
in the longitudinal direction, some directed flow can be generated. 
In Figs. \ref{fig:v1timeht} and \ref{fig:v1time} is shown the 
development of the asymmetric flow in the $x$ direction at different times.
The average velocity in the $x$ direction is calculated for a given 
space-time rapidity and time
\begin{equation}
\langle  v_x \rangle= \frac{\int dx dy v_x \gamma \epsilon}{\int dx dy 
\gamma \epsilon} \ .
\end{equation}
As mentioned above, during the evolution the Bjorken flow $Y=\eta_\parallel$ 
is modified, but the final flow conserves a strong correlation between 
 $Y$ and $\eta_\parallel$ \cite{Bozek:2009ty}. 
Therefore the directed flow as functions 
of space-time rapidity $\eta_\parallel$ as shown in Figs. \ref{fig:v1timeht} 
and \ref{fig:v1time} reflects qualitatively the final flow of particles.

The first observation is that the directed flow is built up for an extended 
time during the expansion, in the first $3$fm/c. 
There is a noticeable difference
between the evolution from the shifted and the tilted initial conditions.
For the shifted initial conditions (Fig. \ref{fig:v1timeht}) 
in the central space-time rapidity region
a positive directed flow develops gradually in the first $1$fm/c. For large 
rapidities a negative flow is generated in the first $3$fm/c. The directed
 flow changes sign around $\eta_\parallel=2$, a behavior very different 
from the one 
observed in the experiment. 
The appearance of the wiggle in the dependence
of $v_1$ on pseudorapidity 
is a consequence of the form of the initial profile of the pressure
(Fig. \ref{fig:indens1}). The gradient of the pressure has a
significant deflection  from the transverse direction only
 in the forward/backward rapidity regions.
On the other hand, the tilted initial condition has a smooth tilted gradient 
of the pressure that gives a negative directed flow in a broad
 range of space-time rapidities (Fig. \ref{fig:v1time}). 
The anti-flow increase for the first $3$fm/c. The negative directed flow
 of the fluid, increasing with rapidity,  leads to a similar pattern in the 
directed flow of the final particles.

\begin{figure}
\includegraphics[width=.5\textwidth]{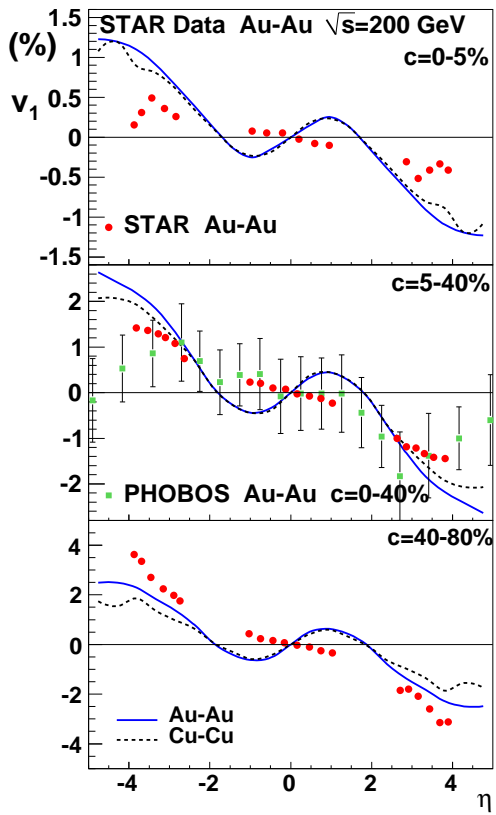}
\caption{(Color online) Directed flow at
 different centralities from hydrodynamic 
calculations with shifted initial conditions (Eq. \ref{eq:nwdens}) for
 Au-Au and Cu-Cu collisions
(solid and dashed lines).  Experimental 
data are from the PHOBOS Collaboration
 \cite{Back:2005pc} for $c=0-40$\%, and from the STAR Collaboration 
\cite{Abelev:2008jga} for the three other centrality classes.}
\label{fig:v1}
\end{figure}

\begin{figure}
\includegraphics[width=.5\textwidth]{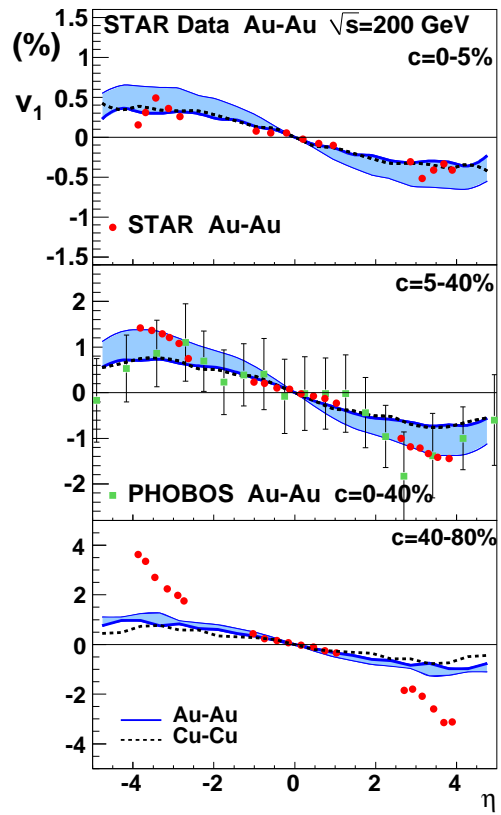}
\caption{(Color online) Directed flow in Au-Au (thick solid lines) and Cu-Cu
 (dashed lines)  collisions at different centralities from tilted initial 
conditions (Eq. \ref{eq:tilt}), compared to experiment 
\cite{Back:2005pc,Abelev:2008jga}. The shaded band  between the thin 
and thick lines represents the increase of the magnitude 
of the flow if Eq. (\ref{eq:tilt2}) is used for the initial density,
including also asymmetric 
contributions from binary collisions.}
\label{fig:v1rad}
\end{figure}

The hydrodynamic evolution is continued until  freeze-out, 
where particle emission from the freeze-out hypersurface 
takes place. Statistical emission and  resonance decays are performed 
using 
the event generator THERMINATOR \cite{Kisiel:2005hn}. 
In Figs. \ref{fig:v1}
 and \ref{fig:v1rad} are shown the results for three representative
centralities, for 
which experimental data for Au-Au collisions have been published. Other 
experimental data show that in Cu-Cu interactions  almost the same directed flow
 is generated as in the larger system if the centrality is chosen to 
be the same \cite{Abelev:2008jga}.
The directed flow for the shifted initial 
conditions shown in Fig. \ref{fig:v1} 
has an incorrect 
dependence on  pseudorapidity. For all the three centrality classes 
the flow is positive in the central rapidity region, and switches 
to anti-flow in the very forward and backward regions. On the same plots 
are shown the results for Cu-Cu collisions in similar centrality classes.
 The flow is similar as for the larger system, and hence different 
than observed experimentally. The similarity between the flow in 
Au-Au and Cu-Cu systems reflects the similarity in the initial density profiles 
(Fig. \ref{fig:indens1}).

In Fig. \ref{fig:v1rad} are shown results for the directed flow of charged particles emitted after a
 hydrodynamic evolution from  the tilted initial fireball 
(Eq. \ref{eq:tilt}). The thick solid lines represent the results for
 Au-Au collisions. The experimental data are reproduced in the central rapidity
region in a satisfactory way. For the semi-peripheral and peripheral   
collisions the 
large anti-flow at large pseudorapidities is not reproduced by the model.
This kinematic region is at the limit of  applicability of the 
hydrodynamic model, assuming the collective expansion of a thermalized fluid.
The directed flow in the fragmentation region can have a different origin, like
 in the 
transport models \cite{Bleicher:2000sx}. On the panels in Fig. 
\ref{fig:v1rad} are shown the 
results for the Cu-Cu system in the same centrality classes. Again, as 
expected from the similarity of the tilt of the source in the two systems 
(Fig. \ref{fig:indens2}), the final directed flows for the two systems 
almost overlap.
To test some of the uncertainty in the choice of the initial conditions 
for the hydrodynamic evolution, we take as the initial density  an ansatz
 where also the contribution from binary collisions is asymmetric
\begin{eqnarray}
\epsilon(\tau_0)&=& 2 \epsilon_0 
 \frac{N_+f_+(\eta_\parallel)+ N_-f_-(\eta_\parallel)}{N_++N_-}\nonumber \\
 && \left[
(1-\alpha) (N_++N_-) 
+  2 \alpha N_{bin} \right] /N_0 \ .
\label{eq:tilt2}
\end{eqnarray}
The tilt is stronger in this case and the magnitude of the directed flow 
bigger (thin solid lines in Fig. \ref{fig:v1rad}). The shaded bands in the
 figure represent the uncertainty of the model related to this assumption. 
This uncertainty is the largest among those that we tested.

\begin{figure}
\includegraphics[width=.5\textwidth]{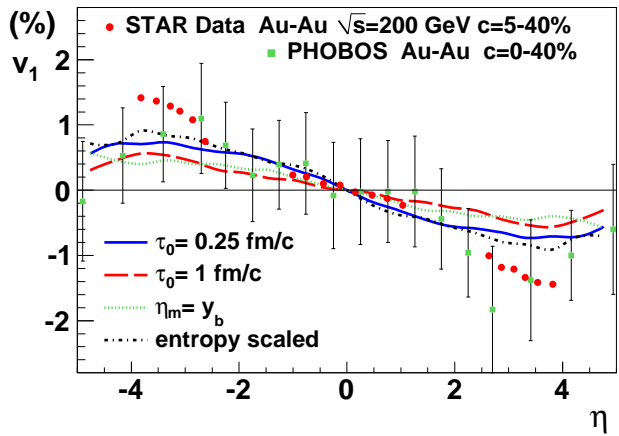}
\caption{(Color online) Directed flow coefficient as function of 
pseudorapidity for tilted initial conditions (Eq. \ref{eq:tilt}) in 
Au-Au collisions. 
The solid and dashed lines represent the calculations with initial 
times $\tau_0=0.25$ and $1$fm/c respectively and with an initial profile $f_F$ 
with $\eta_m=y_b-2$, the dotted line is for $\tau_0=0.25$fm/c but
 $\eta_m=y_b$, and the dashed-dotted line represents the result of a
calculation where the initial entropy density is proportional to 
the density of participants from the 
Glauber Model. Data are from Refs. \cite{Back:2005pc,Abelev:2008jga}.}
\label{fig:v1por}
\end{figure}

In the following we study the effect of other details of the model on the 
final directed flow. First, a different initial time for the hydrodynamic 
evolution is chosen $\tau_0=1$fm/c. The initial time  is given by  the 
rate of the initial thermalization processes, that lead to the formation
of a dense, almost perfect fluid. The precise  mechanisms of thermalization 
are not known and it is instructive to test the influence of the value of 
the thermalization time on the build up of the directed flow. In Fig.
 \ref{fig:v1por} is shown the result for the centrality class $5-40$\%.
The directed flow is reduced when compared to the previous calculation using 
$\tau_0=0.25$fm/c. The fluid that starts to expand at $\tau_0=1$fm/c 
has a smaller 
pressure.
In fact, the experimental points for central rapidities lie
 between the two lines representing the two calculations starting at different 
thermalization times.

Another effect worth to be studied is the dependence on the details of the 
form-factor $f_\pm(\eta_\parallel)$, describing the emission from a single
 participant nucleon. The range of forward-backward correlations is set by the 
parameter $\eta_m$. We make another calculation using $\eta_m=y_b$. With
this choice, the emission from a forward going participant nucleon decreases 
when going from rapidity $y_b$ to zero at the rapidity $-y_b$. The initial tilt 
of the source is smaller than for the choice $\eta_m=y_b-2$, used before.
The effect of this change on the directed flow is small and the results 
are close to the experimental data at central rapidities 
(dotted line in Fig. \ref{fig:v1por}).

A third effect, that can be tested, is the assumption that the initial energy 
density is proportional to the density of wounded nucleons (binary collisions)
as in Eq. (\ref{eq:tilt}). Assuming that this proportionality applies to 
the entropy density $s$ instead, we have
\begin{eqnarray}
s(\tau_0)&=& s_0 \left[
 \left( N_+(x,y)f_+(\eta_\parallel)+ N_-(x,y)f_-(\eta_\parallel)\right)
(1-\alpha) \right.\nonumber \\
&+& \left. 2 \alpha N_{bin}(x,y) f(\eta_\parallel)\right] /N_0 \ .
\label{eq:tilts}
\end{eqnarray}
The directed flow is very similar to the one obtained using the 
energy density initial profile (\ref{eq:tilt}) (dashed-dotted 
line in Fig. \ref{fig:v1por}). We notice that some of the 
details of the initial conditions can influence the final directed flow. This 
situation resembles the conclusion from the elliptic flow studies, where 
a crucial
 ingredient is the initial eccentricity of the fireball.

\section{Summary}

We study the formation of the directed flow 
in the $3+1$D hydrodynamic expansion of
 the fireball created in heavy-ion collisions at the highest energies at RHIC.
The directed flow of charged particles has been measured as function
 of pseudorapidity, finding a substantial negative flow 
\cite{Back:2005pc,Adams:2005ca,Abelev:2008jga}. We use two 
different initial conditions for the evolution. The first one 
(Eq. \ref{eq:nwdens}) is quite commonly
used in hydrodynamic model calculations \cite{Hirano:2002ds}. It incorporates a 
shift of the densities in the initial fireball, due to the local imbalance 
of the longitudinal momentum.  Initial densities of the second type are 
constructed as a sum of contributions from forward and backward going
 participants (Eq. \ref{eq:tilt}). The asymmetry in the emission from 
individual participants leads to a tilt of the source.
Model calculations, incorporating a hydrodynamic expansion stage, 
particle emission at freeze-out, and resonance decays, 
indicate that the second type of initial conditions can reproduce the 
sign and the magnitude of the observed directed flow  at central rapidities.

The large negative flow close to the fragmentation regions is of 
different origin and  cannot be 
described in our calculation. Deviations from the Bjorken 
flow in the initial conditions
 \cite{Snellings:1999bt,*Becattini:2007sr} could influence the results. 
In particular, they  could lead to  a different flow pattern for baryons 
and for  the bulk of the matter, if  baryons do not follow the 
Bjorken flow. The directed flow is expected to decrease with the collision
energy  for two reasons. With increasing kinematic range 
of rapidities the tilt of the source goes down, also the contribution of
 binary collisions to the Glauber Model density is believed to increase 
with the energy, with a similar consequence. Finally let us note, 
that our calculation reproduces the 
experimentally observed similarity of the flow in Au-Au and Cu-Cu collisions
at the same centrality.

\begin{acknowledgments}
The authors are very grateful to
Jean-Yves Ollitrault for  correspondence and discussions that  have been 
 of vital importance in 
finding errors in early studies and in reaching the correct conclusions. 
The work is supported  by 
Polish Ministry of Science and Higher Education under
grants N202~034~32/0918 and N~N202~263438.
\end{acknowledgments}

\bibliography{../hydr}

\end{document}